\documentclass[twocolumn,showpacs,preprintnumbers,amsmath,amssymb,floatfix,prl]{revtex4}
\usepackage{epsfig}

\begin{document}

\title{Formation of Nanopillar Arrays in Ultrathin Viscous Films:\\
The Critical Role of Thermocapillary Stresses}
\author{Mathias Dietzel and Sandra M. Troian
\footnote{Corresponding author: stroian@caltech.edu.}}
\affiliation{California Institute of Technology,\\
1200 E. California Blvd. MC 128-95, Pasadena, CA 91125}

\begin{abstract}
Experiments by several groups during the past decade have shown that a molten polymer nanofilm subject to a large transverse thermal gradient undergoes spontaneous formation of periodic nanopillar arrays. The prevailing explanation is that coherent reflections of acoustic phonons within the film cause a periodic modulation of the radiation pressure which enhances pillar growth. By exploring a deformational instability of particular relevance to nanofilms, we demonstrate that thermocapillary forces play a crucial role in the formation process. Analytic and numerical predictions show good agreement with the pillar spacings obtained in experiment. Simulations of the interface equation further determine the rate of pillar growth of importance to technological applications.

\end{abstract}


\maketitle


Fabrication of high resolution, large area arrays for micro-optic,
photonic and optoelectronic devices
relies heavily on photolithographic patterning
by projection of UV light.
This process is inherently slow and costly due to multiple step-and-repeat procedures required for deposition,
exposure and removal of photoresist layers for construction of 3D
components. Harsh developer and etching solutions also
imbue structures with significant surface roughness, which ultimately
limits performance due to scattering losses. Photoresist masks must also be UV compatible, which restricts
the assortment of materials used in processing.
Alternative patterning techniques based on
''resistless lithography'' may usher
a new era in lithographic patterning in which structure formation is obtained by directed deposition or mechanical embossing of material as with microcontact printing, micromolding, microembossing and nanoimprinting \cite{Guo:AdvMat2007,Rogers:ChemRev2007}. An even newer approach for emergent technologies relies on film patterning by hydrodynamic instabilities in nanoscale films as with templated dewetting \cite{Mayr:JAP2008}. The use of fluid instabilities for controlled formation of large area patterning provides an interesting route for non-contact fabrication of periodic microscale and nanoscale arrays.

Several groups pursuing this approach have investigated the spontaneous formation of pillar arrays in molten polymer nanofilms
subject to a large transverse thermal gradient \cite{ChouGuo:APL1999,ChouZhuang:JVacSciTech1999,
SchafferSteiner:EurophysLett2002,SchafferSteiner:Macromolecules2003,SchafferSteiner:AdvMat2003,
PengHan:Polymer2004}. The prevailing explanation
\cite{SchafferSteiner:Macromolecules2003,SchafferSteiner:AdvMat2003} is that coherent reflections of acoustic phonons (AP) within the film cause periodic modulation of the radiation pressure, similar to an acoustic Casimir interaction force \cite{Steiner:PRL2004}. This normal stress opposes capillary forces and enhances the growth of protrusions. Such a process, however, requires coherent phonon propagation within a molten amorphous polymer film and an average phonon mean free path at least as large as the film thickness. Experiments have shown that solid nanoscale polymer films at temperatures $-193 ^o \textrm{C} \leq T \leq 27 ^o \textrm{C}$ can support phonon attenuation lengths of about  $\textsf{O}(10^1-10^2)$ nm at frequencies in the 100 GHz range \cite{MorathMaris:PhysRevB96}. However, such long propagation lengths  have never been demonstrated and are considered unlikely in amorphous molten films (glass transition temperature 100 - 120 $^o \textrm{C}$).
\vspace{-0.2in}
\begin{figure}[tbhp]
\includegraphics[width=9.0cm]{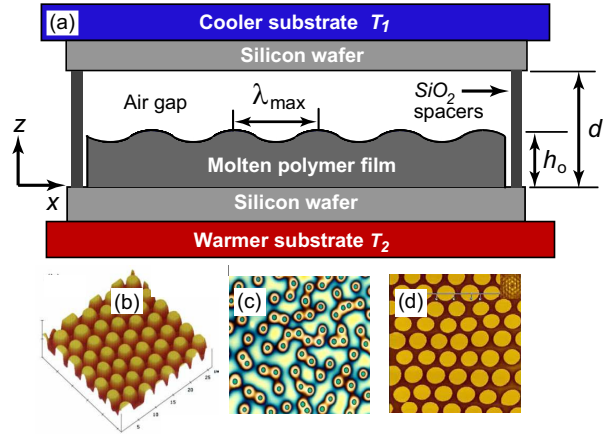}
\caption{\footnotesize (a) Experimental setup for formation of nanopillar
arrays. (b) AFM image of PMMA pillars
\cite{ChouZhuang:JVacSciTech1999}: $d=260$ nm, $h_o=95$ nm,
$\lambda_{\textsf{max}}=3.4 \mu$m, $\Delta T$ unknown. (c) Optical
micrograph of PS pillars \cite{Schaffer_phd01}: $d=345$ nm,
$h_o=100$ nm, $\lambda_{\textsf{max}}=4.5 \mu$m, $\Delta T = 46$ $^o \textrm{C}$.
(d) AFM image of PMMA pillars \cite{PengHan:Polymer2004}: $d=163$ nm,
$h_o=100$ nm, $\lambda_{\textsf{max}}=6.5 \mu$m, $\Delta T = 10$ $^o \textrm{C}$
\normalsize.} \label{apparatus}
\end{figure}
By investigating an unexplored limit of interfacial instability easily accessible to nanoscale films, we demonstrate that thermocapillary (TC) forces play a crucial if not dominant role in this formation process. According to this mechanism, perturbations in film thickness generate periodic disturbances in the surface temperature which lead to periodic modulation of the interfacial thermocapillary stress. These tangential stresses cause the polymer melt to flow toward cooler regions, thereby driving protrusions toward a cooler target substrate. These elongated pillars are observed to solidify rapidly in place as soon as the driving force is removed. Nanostructures fabricated in such non-contact fashion from the molten state are expected to exhibit superior optical performance due to the specularly smooth interface obtained upon solidification. Investigation of this new regime of instability is therefore pertinent to fundamental studies of thermocapillary flow as well as technological innovations for non-contact resistless lithography.

A schematic diagram of the typical experimental setup is shown in
Fig.~\ref{apparatus}(a). Polymers like polystyrene (PS) or poly(methyl
methacrylate) (PMMA) are dissolved in solvent and spun cast onto a silicon wafer to a thickness $80 \:\textrm{nm} \lesssim h_o \lesssim 130 \: \textrm{nm}$. The coated wafer is overlaid by a second wafer and vertical spacers are used to maintain a separation distance $d > h_o$ where
$100 \: \textrm{nm} \lesssim d \lesssim 600 \: \textrm{nm}$. Ratios of $d/h_o$ in experiment range from approximately 2-8. The
bottom wafer is placed on a hot substrate held at temperature
$130 ^o \textrm{C} \lesssim T_2 \lesssim 170 ^o \textrm{C}$; the
polymer free surface is cooled from above by proximity to the cold
substrate held at $T_1 < T_2$. Both temperatures are maintained above the polymer glass transition to ensure a molten film during the formation process. Despite the fact that $10 ^o \textrm{C} \lesssim T_2 - T_1 \lesssim 55 ^o \textrm{C}$ is not large, the small gap size establishes a very large transverse gradient
$\Delta T/d \sim 10^6 - 10^8$ $^o \textrm{C}$/cm. In the experiments under study, films were subjected to a thermal gradient overnight and then quenched to room temperature to solidify the structures formed. The top wafer was then removed and optical microscopy or atomic force microscope (AFM) images obtained, which revealed the  patterns shown in Fig.~\ref{apparatus} (b)-(d).

It is well known that much thicker liquid films (cm to mm) subject to
much smaller thermal gradients can develop periodic cellular patterns through Rayleigh-B\'{e}nard (RB) or B\'{e}nard-Marangoni (BM) instability \cite{Probstein1994}. These instabilities, however, generate shallow corrugations not needle-like protrusions. Onset of instability requires that the critical Rayleigh number $Ra_c$ for buoyancy driven flow (which scales as $h^4_o$) or the critical Marangoni number $Ma_c$ for thermocapillary flow (which scales as $h^2_o$) exceeds 660 - 1700 or 50-80, respectively, depending on boundary conditions.
For the nanoscale films shown in Fig.~\ref{apparatus}, $Ra \approx
10^{-16}$ and $Ma \approx 10^{-8}$, ruling out these mechanisms as possible causes for pillar formation.
A lesser known deformational instability leading to more pronounced dry spots or elevations \cite{Smith:JFM1966}
has recently been observed in microscale films ($50 \lesssim h_o \lesssim 250 \mu \textrm{m}$) \cite{Bestehorn:EuroPhysJB2003,VanHook:JFM1997,VanHook:PRL1995} in which the thermocapillary stress is counterbalanced by capillary and
gravitational forces. Onset of instability requires that the inverse dynamic Bond number
$D_c=\gamma_T \Delta T_\textrm{film} /\rho g {h_o}^2 \geq 2/3(1+F)^{-1}$, where $\rho$ is the liquid density,
$\gamma_T \equiv |d\gamma/dT|$, $\gamma$ is the liquid surface tension, $\Delta T_\textrm{film}$ is the temperature drop across the liquid
layer, $F=(1-\kappa)/(D + \kappa -1)$ is an order one constant, $D=d/h_o$, and $\kappa=k_\textrm{air}/k_\textrm{liq}$ is the ratio of thermal conductivities.
Parameter values for the experiments in Fig.~\ref{apparatus}
reveal that $D_c \gtrsim \textsf{O}(10^7)$ and $G \sim \textsf{O}(10^{-14})$,
far beyond regimes of instability previously
investigated in which $D_c \sim \textsf{O}(10^{-1} - 1)$ and $G \sim \textsf{O}(10^{-1} - 10^2)$.

In this paper, we investigate unexplored consequences of this deformational instability for nanoscale films subject to a very large transverse gradient.  The governing geometric and dynamic parameter ranges are constrained by $\epsilon^2 = (h_o/\lambda^\textrm{TC}_{\textrm{max}})^2 \ll 1$, $\epsilon Re \rightarrow 0$, $\epsilon Re Pr \rightarrow 0$ and $G \rightarrow 0$ i.e. gravitational stabilization is absent.
The lateral scale for the slender gap ratio, $\epsilon$, is set by  $\lambda^\textrm{TC}_{\textrm{max}}$, the wavelength of the maximally unstable mode (i.e. fastest growing mode) obtained from linear stability analysis, which in experiment corresponds to the
average pillar spacing. Here,
$Pr = \nu/ \alpha$ is the Prandtl number,
$G = g {h_o}^3/ \nu \alpha $ is the Galileo number, $Re = u_c h_o/\nu$ is the Reynolds number, $u_c$ is the characteristic
lateral flow speed set by thermocapillary forces, $g$ is the gravitational acceleration constant, $\nu$ is the
polymer kinematic viscosity and $\alpha$ is
the polymer thermal diffusivity, both evaluated at the temperature $T_2$.  These constraints establish the slender gap approximation for momentum and thermal
transfer \cite{Leal07} in the limit where there is no hydrostatic restoring force.

Within this approximation, the differential equations for momentum and energy conservation decouple completely \cite{OronDavis:RevModPhys1997}. The energy equation reduces to a 1D steady state process for thermal conduction across an air/polymer bilayer with an internal undulatory interface $H(X,Y,\tau)$; variation of the interfacial temperature with time arises solely through displacement of this interface.
The relevant variables are normalized according to
$(X,Y)=(x/\lambda^\textrm{TC}_{\textrm{max}},y/\lambda^\textrm{TC}_{\textrm{max}})$, $Z=z/h_o$, $H=h/h_o$, $\Theta=(T-T_1)/\Delta T$ where $\Delta T= T_2 - T_1$, $\tau = u_c t/L$, $P=\epsilon h_o p/\eta u_c$
and $\Gamma = \epsilon \gamma/\eta u_c$ where
$t$, $p$ and $\eta=\eta(T_2)$ denote real time, pressure and melt viscosity.
The continuity equation for incompressible flow yields the relevant
scaling for the velocity fields; namely $\vec{U}= (U,V,W) =
(u/u_c,v/u_c,w/\epsilon u_c)$. The boundary conditions for the velocity and stress fields are the usual
no-slip condition, impenetrability at the solid surface $Z=0$, and a jump in the normal and tangential stresses at $Z=H(X,Y,\tau)$. In the slender gap approximation, these stress conditions reduce to $P_{Z=H}=- \overline{Ca}\:^{-1}\nabla_s^2 H$ and $\nabla_s \Gamma
=-\overline{Ma} \nabla_s \Theta_{Z=H}$ where $\nabla_s \Theta_{Z=H} = - \kappa D \nabla_{\parallel}H /[D + (\kappa -1)H]^{-2}$,
the thermal Marangoni number $\overline{Ma}=\epsilon \gamma_T \Delta T/(\eta u_c)$,
the surface gradient operator $\nabla_s \rightarrow \nabla_{\parallel} = (\partial/\partial X,\partial/\partial Y)$
\cite{Dean98}, and $\overline{Ca}=\eta u_c/(\gamma \epsilon^3)$.
The kinematic boundary condition, $dH/d\tau=W_{Z=H}$, is then re-expressed
to yield the 4th-order, non-linear equation for the
evolution of the air/polymer interface \cite{Leal07}:
\begin{equation}
\frac{\partial H}{\partial \tau} + \nabla_{\parallel} \cdot
\left( \frac{\kappa D \overline{Ma} H^2}{2[D + (\kappa -1)H]^2} \nabla_{\parallel} H  + \frac{H^3}{3\overline{Ca}} \nabla^3_{\parallel}H \right)=0.
\label{TCinterface}
\end{equation}
In our simulations, $u_c$ was set by the choice that the interfacial thermocapillary stress and $\nabla_{\parallel}H$  equal unity i.e. $(\partial U/\partial Z)_{Z=H} = \partial \Gamma /\partial X = 1$, such that $u_c=(4 \pi)^2 \gamma \epsilon^3 /3 \eta$ \cite{footnoteA}.
With this choice, $\overline{Ca} = (4\pi)^2/3$ and
$\overline{Ma}= 3 \gamma_T \Delta T/[(4 \pi)^2 \gamma \epsilon^2]= [D + (\kappa - 1)]^2/(\kappa D) \sim \textsf{O}(10^{-1} - 10^{1})$.

For the parameters values pertinent to experiment, spatial gradients in $H$ for times $\tau \lesssim 1$ are $\textsf{O}(1)$ ; consequently, the first term
in Eq. (\ref{TCinterface}) is $\textsf{O}(1)$, the second term is $\textsf{O}(10^{0} - 10^{1})$, and the third term is $\textsf{O}(10^{-2})$. The destabilizing thermocapillary term is therefore $\textsf{O}(10^2 - 10^3)$ larger than the stabilizing capillary term, which is the reason that pillar-like structures can form in this system. A similar derivation leading to Eq. (\ref{TCinterface}) yields the corresponding equation for the acoustic phonon model \cite{SchafferSteiner:Macromolecules2003}:
\begin{equation}
\frac{\partial H}{\partial \tau} + \nabla_{\parallel} \cdot
\left( \frac{\bar{Q}(1 - \kappa ) H^3}{3\overline{Ca} [D + (\kappa - 1 )H]^2} \nabla_{\parallel} H + \frac{H^3}{3\overline{Ca}} \nabla^3_{\parallel} H \right)=0,
\label{APinterface}
\end{equation}
where $\bar{Q} = 2Q k_a \Delta T/(u_p \gamma \epsilon^2)$,
$Q$ is a phenomenological reflectivity coefficient, $u_p$ denotes the speed of sound in polymer, and $\epsilon=h_o/\lambda^\textrm{AP}_{\textrm{max}}$.
The liquid flux in Eq. (\ref{TCinterface}) due to shear flow by thermocapillary forces is instead replaced in Eq. (\ref{APinterface}) by a pressure driven flow due to acoustic phonons reflecting from the film interfaces.

Conventional linear stability analysis yields the
(dimensional) wavelength for the maximally unstable mode corresponding to these different models, namely
\begin{equation}
\lambda^\textrm{TC}_{\textrm{max}}= 2 \pi h_o \sqrt{\frac{4
\gamma h_o}{3 \: \kappa \:d \:\gamma_T \:\Delta T }}\left[\frac{d}{h_o} + \kappa -
1\right]
\label{tcwavelength}
\end{equation}
versus
\begin{equation}
\lambda^\textrm{AP}_{\textrm{max}} =
2 \pi h_o \sqrt{\frac{\gamma \: u_p}{Q (1-\kappa) k_a \Delta T}}\left[\frac{d}{h_o} + \kappa - 1\right].
\label{apwavelength}
\end{equation}
These characteristic length scales can be
directly compared to the pillar spacings observed in experiment. Linear stability also yields the dimensionless cutoff wave number $K_c$ below which all modes are unstable (discussed further in the summary) . For the TC model, $K^{TC}_c = (3/2 \overline{Ma}\:\overline{Ca}\kappa D )^{1/2}/[D + (\kappa - 1)].$
\begin{figure}[tbhp]
\includegraphics[width=8.5cm]{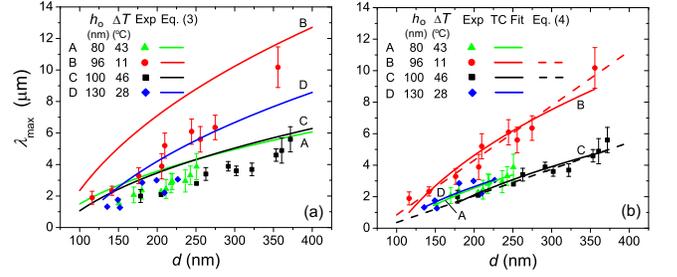}
\caption{\footnotesize (a) Experimental data from
Refs.~\cite{SchafferSteiner:EurophysLett2002,
SchafferSteiner:Macromolecules2003,SchafferSteiner:AdvMat2003,Schaffer_phd01}.
Material constants, evaluated at the temperature $T_2 = 170$ $^o$C,
were obtained from
Refs.~\cite{Mark:AIPPress1996,Moreira:JAPS2001,Masson:PRE2002,Lide:CRCPubComp1992}.
(b) Fitting coefficients are of the form [Expt A-D, $C_1~(10^3 \mu m)^{0.5}$,
$C_2~(0.1 \mu m)^{1.5}$]: [A, 0.353, -34.7], [B, 0.650, -64.6], [C,
0.379, -46.0], [D, 0.340, -30.7]. For the AP model, $Q=6.2$ and $u_p$
= 1850 m/s. \normalsize} \label{lambda}
\end{figure}
Solutions of Eq.~(\ref{tcwavelength}) shown in Fig.~\ref{lambda}(a) contain no adjustable parameters. While the functional dependence of Eq.~(\ref{tcwavelength}) on $d$ is in good agreement with experiment, the TC model systematically over predicts the average pillar spacing, in some cases by as much as 35 - 40\%. To further test the dependence of
$\lambda^\textrm{TC}_{\textrm{max}}$ on $d$, we performed a least-squares fit of the experimental data to the form of Eq.~(\ref{tcwavelength}), namely $C_1 d^{1/2} + C_2 d^{-1/2}$, as shown in Fig.~\ref{lambda}(b).

\begin{figure}[tbhp]
\includegraphics[width=8.5cm]{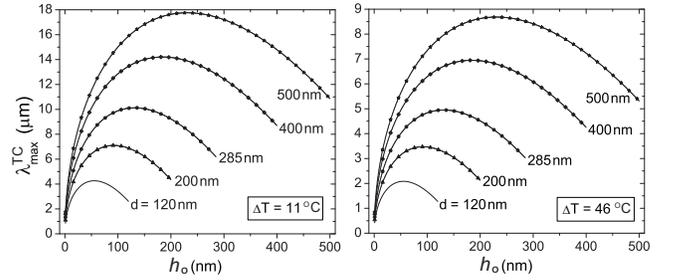}
\caption{\footnotesize Solutions of Eq.~\ref{tcwavelength}
for $\Delta T = 11$ $^o$C and $46$ $^o$C. \normalsize}
\label{lamb_sensitivity}
\end{figure}

Superimposed on these curves is also a least-squares fit to Eq.~(\ref{apwavelength}) with fitting parameters $Q=6.2$ and $u_p$ = 1850 m/s \cite{SchafferSteiner:Macromolecules2003}.
We have determined that several experimental factors may contribute to the offset observed in Fig.~\ref{lambda}(a); here we focus on a key issue involving the measured values of $h_o$ in the literature. Spin cast polymer films are prone to significant solvent retention \cite{Garcia:CollPoly2007,Perlich:Macromol2009}, typically eliminated by post annealing for several hours in a vacuum oven at elevated temperatures. Significant shrinkage in film thickness has been reported based on ambient vapor pressure, time and temperature of the bake. In the experiments of Ref.\cite{SchafferSteiner:Macromolecules2003,SchafferSteiner:AdvMat2003,PengHan:Polymer2004,Schaffer_phd01}, there are no reports of film annealing following spin casting, which would lead to overestimates of $h_o$. Fig.~\ref{lamb_sensitivity} indicates the strong dependence of $\lambda^\textrm{TC}_{\textrm{max}}$ on $h_o$ and $d$ and a sharp drop in the predicted wavelength for small values of $\Delta T$ and $h_o$. While smaller values of $h_o$ due to film shrinkage leads to very good agreement, we note that solvent evaporation is assumed to occur prior to insertion in the experimental assembly and therefore plays no role in pillar formation.

\begin{figure}[tbhp]
\includegraphics[width=8.5cm]{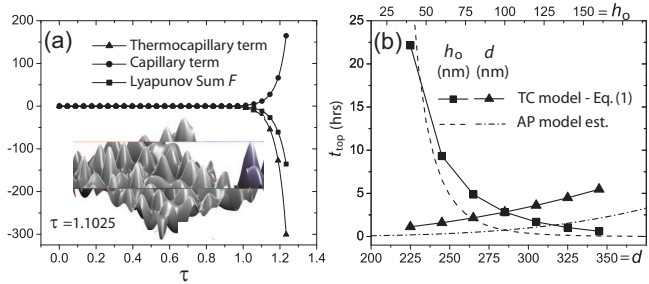}
\caption{\footnotesize (a) Evolution of the Lyapunov free energy for the
case $h_o$=100 nm, $d$=285
nm, $\Delta T=46 ^o \textrm{C}$.
Real time conversion $t=2.28 \tau$ hrs. Inset: Plot of
$H(X,Y,\tau=1.1025)$ from finite element simulations of Eq. (\ref{TCinterface}). (b) Time
required, $t_\textrm{top}$, for fastest growing pillars to contact
upper wafer for parameter values given in (a). For variation with
$h_o$, $d$=285 nm; for variation with $d$, $h_o$=100 nm. \normalsize} \label{Lyapunov}
\end{figure}

To better understand this instability, we examined the Lyapunov free energy \cite{OronRosenau:JPhysIIFrance1992,Oron:PoF2000} $F=\int \mathfrak{L} ~ dX dY$  for the evolving film where
\begin{equation}
\mathfrak{L}\!=\!(\nabla_{\parallel} H)^2 \!- \!\frac{3 \kappa \overline{Ma}\: \overline{Ca}}{D} \left[ H \textrm{ln} \left (\frac{H}{1 + \chi H} \right) \!+\!\textrm{ln} (1+\chi) \right].
\end{equation}
and $\chi = (\kappa - 1)/D$. The 1st and 2nd terms on the right represent the capillary and thermocapillary contributions, respectively. Finite element simulations with periodic boundary conditions and 2nd order Lagrangian shape functions for spatial discretization of the film height were conducted. As shown in Fig.~\ref{Lyapunov}(a), once $\tau$ exceeds unity, film evolution enters the nonlinear regime, as indicated by the steep drop in free energy due to the overwhelming influence of thermocapillary forces. The terminating points correspond to contact with the upper plate.
This instability is non-saturating (i.e. no steady state solution in contrast to RB or BM instabilities); pillars will simply continue to grow until contact with the upper plate. Shown in Fig.~\ref{Lyapunov}(b) are predictions of the time required for the fastest growing pillars to contact the cooler plate. The TC curve was obtained from Eq.(\ref{TCinterface}); the AP curve is an estimate based on the growth rate corresponding to Eq. (\ref{apwavelength}) from linear stability analysis. Our estimates in Fig.~\ref{Lyapunov}(b) also indicate that the fastest growing nanopillars contact the upper substrate within a few hours of formation, while the experiments typically lasted overnight.  Filaments bridging both substrates  might then undergo capillary or even themocapillary migration toward narrower gaps within the slight wedge geometry used in experiment (tilt $\approx 1 \mu$m per cm according to \cite{Schaffer_phd01}). This secondary effect would also lead to better agreement due to smaller measured values of $\lambda^\textrm{TC}_{\textrm{max}}$. Much larger values of $D=d/h_o$ were also investigated numerically. Evolution into the nonlinear regime corresponding to much larger pillar amplitudes generates in-plane hexagonal ordering, in agreement with experimental observations\cite{Dietzel:JAP2009}.

In summary, we have shown that thermocapillary stresses play a crucial if not dominant role in the formation of pillar arrays in molten nanofilms subject to a large transverse thermal gradient. The parameter range investigated  corresponds to an unexplored limit of deformational instability in which destabilizing thermocapillary forces far outweigh stabilization by capillary or gravitational forces. The predominance of thermocapillary effects allows the formation of elongated nanostructures. Our analysis indicates that any Newtonian liquid subject to even smaller thermal gradients will undergo  pillar formation for wavenumber disturbances $K < K_c$. However, technological demand for high resolution optical or other large area arrays with very small feature sizes suggests use of very large thermal gradients, smaller gap widths, and smaller film thicknesses $h_o$.

We gratefully acknowledge support from the NSF-ECCS Division and early discussions with Dr. A. Darhuber at Princeton University.


\end{document}